\documentclass[conference]{IEEEtran}
\IEEEoverridecommandlockouts
\usepackage[dvipdfmx]{graphicx}
\usepackage{color}
\usepackage{colortbl}
\usepackage{float}

\usepackage{array}
\usepackage{multirow}
\usepackage{subcaption}
\usepackage{threeparttable}

\usepackage{gensymb}
\usepackage{amsmath,amssymb,amsfonts}
\usepackage{algorithmic}
\usepackage{comment}
\newcommand{\tablenl}[1]{\begin{tabular}{c} #1 \end{tabular}}
\newcommand{\mm}{mm\textsuperscript{2}}
\newcommand{\cer}{\,\degree\text{C}}

\usepackage[backend=bibtex,style=ieee]{biblatex}
\usepackage{hyperref}
\DefineBibliographyStrings{english}{
  url         = \adddot\addspace URL ,
}

\usepackage[whole]{bxcjkjatype}
\usepackage{siunitx}

\usepackage{braket}
\usepackage{textcomp}
\usepackage{paralist}
\usepackage[a4paper, total={184mm,239mm}]{geometry}
\def\BibTeX{{\rm B\kern-.05em{\sc i\kern-.025em b}\kern-.08em
    T\kern-.1667em\lower.7ex\hbox{E}\kern-.125emX}}
\begin{document}

%baseline stretch

\title{Lookup Table-based Multiplication-free All-digital DNN Accelerator Featuring Self-Synchronous Pipeline Accumulation}

 \author{\IEEEauthorblockN{Hiroto Tagata}
 \IEEEauthorblockA{\textit{Graduate School of Informatics} \\
 \textit{Kyoto University}\\
 Kyoto, Japan \\
 htagata@easter.kuee.kyoto-u.ac.jp}
 \and
 \IEEEauthorblockN{Takashi Sato}
 \IEEEauthorblockA{\textit{Graduate School of Informatics} \\
 \textit{Kyoto University}\\
 Kyoto, Japan \\
 takashi@i.kyoto-u.ac.jp}
 \and
 \IEEEauthorblockN{Hiromitsu Awano}
\IEEEauthorblockA{\textit{Graduate School of Informatics} \\
 \textit{Kyoto University}\\
 Kyoto, Japan \\
 awano@i.kyoto-u.ac.jp}
% \and
% \and
% \and
 }

\maketitle

\begin{abstract}
Deep neural networks (DNNs) have been widely applied in our society, yet reducing power consumption due to large-scale matrix computations remains a critical challenge. MADDNESS is a known approach to improving energy efficiency by substituting matrix multiplication with table lookup operations. Previous research has employed large analog computing circuits to convert inputs into LUT addresses, which presents challenges to area efficiency and computational accuracy. This paper proposes a novel MADDNESS-based all-digital accelerator featuring a self-synchronous pipeline accumulator, resulting in a compact, energy-efficient, and PVT-invariant computation. Post-layout simulation using a commercial 22nm process showed that 2.5$\times$ higher energy efficiency (174 TOPS/W) and 5$\times$ higher area efficiency (2.01 TOPS/{\mm}) can be achieved compared to the conventional accelerator.
\end{abstract}

\begin{IEEEkeywords}
Approximate Matrix Multiplication, Computing-in-Memory, Convolutional Neural Network, SRAM, Binary Decision Tree.
\end{IEEEkeywords}

\section{Introduction}
\label{chap:introduction}

%背景の引用を増やす

Deep neural networks (DNNs) are now widely applied in various fields, such as pattern recognition and natural language processing~\cite{cnn2017, resnet2016,dnn2015}. As their performance advances, DNNs continue to expand in parameter size\cite{parameter2022}, resulting in increased inference costs, mainly due to the high computational demands of multiply-and-accumulate (MAC) operations. In particular, the memory access needed to retrieve large amounts of weight parameters during inference is a significant factor in increasing energy consumption and latency~\cite{eyeriss2017}.

Computing-in-Memory (CIM) is a leading approach to reducing memory access costs by integrating computation circuits directly into memory, thereby enhancing inference efficiency~\cite{cim2017}. CIM is categorized into two types: analog and digital. Analog CIM employs voltage, current, and delay for computation, enabling high-speed, energy-efficient MAC operations without relying on advanced process nodes~\cite{c2c2023}. However, its inference accuracy is susceptible to process, voltage, and temperature (PVT) variations, and the increasing demand for multi-bit input/output support has introduced substantial costs associated with DA/AD conversion~\cite{9586103, yoshioka2024}. In contrast, digital CIM offers high precision and robustness against PVT variations, with the potential for enhanced power, performance, and area (PPA) through scaling~\cite{tsmc2024, 7tsram2023}. However, it relies on large multipliers and adders, constraining energy efficiency. In particular, multipliers consume 6 to 31 times more energy and occupy 8 to 25 times more area than adders~\cite{horowitz2014, 9787538}.
%Computing-in-Memory (CIM) is a leading approach to reducing memory access costs by integrating computation circuits directly with memory to enhance inference efficiency. There are two types of CIM: analog CIM and digital CIM. Analog CIM uses voltage, current, and delay for computation to achieve high-speed, energy-efficient MAC operations without the necessity for advanced process nodes~\cite{c2c2023}. However, PVT variations can degrade inference accuracy. Furthermore, the recent trends towards multi-bit I/O support have introduced significant costs associated with DA/AD conversion~\cite{9586103,yoshioka2024}. In contrast, digital CIM provides high precision and is robust against PVT variations, with the potential for improved PPA through scaling~\cite{tsmc2024, 7tsram2023}. However, it requires large multipliers and adders, limiting energy efficiency.
% Nevertheless, it relies on large multipliers and adders, restricting energy efficiency. 
%In particular, multipliers require 6 to 31 times larger energy and 8 to 25 times larger area than adders~\cite{horowitz2014,9787538}.

One solution to reduce the high computational cost of multipliers is approximate matrix multiplication (AMM) using lookup tables (LUT)~\cite{lacc2019, biqgemm2020, maddness2021, pecan2022, lutnet2023}. Especially, MADDNESS (Multiply-ADDitioN-lESS)\cite{maddness2021} employs a product quantization (PQ) technique\cite{pq2011}, whereby the input activation is approximated by prototypes that have been learned prior to the inference. Since the multiplication of the prototypes and known synaptic weight can be computed offline, the precomputed results can be stored in LUT. 
%Consequently, the matrix multiplication can be computed by a low-cost table lookup operation, thereby significantly reducing the computational energy. 
This approach enables matrix multiplication through low-cost table lookup operations, significantly reducing computational energy.
Although PQ introduces some accuracy loss, previous studies report that 16 or more prototypes can achieve accuracy comparable to conventional DNNs~\cite{maddness2021, pecan2022, lutnet2023}.

Nevertheless, the implementation of MADDNESS in hardware presents significant challenges. The computational costs associated with MADDNESS are dominated by the PQ and data readout cost from LUT. However, GPU lacks optimization for these operations, limiting both throughput and energy efficiency ~\cite{mccarter2022lookupsyetneeddeep}. While some MADDNESS accelerators have been proposed, challenges remain: for example, \cite{fuketa2023} uses analog computation in PQ, making it vulnerable to accuracy degradation from PVT variations. Furthermore, the synthesizable accelerator proposed in \cite{stellanera} has high LUT read energy attributed to the redundant layout, reducing energy efficiency.

This paper proposes a MADDNESS-based all-digital accelerator featuring a self-synchronous pipeline accumulator. The proposed circuit employs a data-driven asynchronous design, eliminating the need for a global clock and providing robust PVT variation tolerance. Furthermore, it incorporates a dataflow-based decision tree classifier for the PQ circuit and an area-efficient memory using 10-transistor static random access memory (10T-SRAM), achieving substantial reductions in energy consumption compared to the previous design.

The contributions of this paper are as follows:
\begin{itemize}
    \item This study introduces the first compact, energy-efficient, and PVT-invariant MADDNESS accelerator with a self-synchronous pipeline architecture.
    \item Post-layout simulation using Synopsys HSPICE\textregistered on a commercial 22nm bulk-CMOS process demonstrated 2.5$\times$ enhancement in energy efficiency (174 TOPS/W) and 5$\times$ improvement in area efficiency (2.01 TOPS/\mm) in comparison to the conventional accelerator~\cite{fuketa2023}.
\end{itemize}

The remainder of this paper is organized as follows: Section \ref{chap:related} introduces the MADDNESS algorithm and reviews prior LUT-based accelerators. Section \ref{chap:proposed} details the proposed all-digital MADDNESS accelerator, which incorporates self-synchronous pipeline accumulation. Section \ref{chap:evaluation} presents a performance evaluation based on post-layout simulations in a 22nm process and provides comparisons with conventional accelerators. Finally, Section \ref{chap:conclusion} summarizes this paper.

\section{Background}
\label{chap:related}

MADDNESS\cite{maddness2021} is a technique to reduce neural network inference costs by using approximate matrix multiplication (AMM) based on product quantization (PQ). It improves the inference accuracy after quantization by incorporating codebook generation into the training process of the neural network. This section provides an overview of the fundamental algorithms underlying PQ and MADDNESS, followed by a review of prior research on MADDNESS-based DNN accelerators.

\subsection{Product Quantization}

PQ is a dimensionality reduction technique that divides a high-dimensional vector space into smaller subspaces and quantizes each subspace independently~\cite{pq2011}. To provide a clear illustration of PQ, consider the following example with a $d$-dimensional vector, $\mathbf{x} \in \mathbb{R}^d$. Initially, $\mathbf{x}$ is partitioned into $M$ subvectors, resulting in:
\begin{equation}
    \label{eq:splitx}
    \mathbf{x} = [\mathbf{x}_1, \mathbf{x}_2, \ldots, \mathbf{x}_M], \quad \mathbf{x}_i \in \mathbb{R}^{d/M}
\end{equation}
For each subvector $\mathbf{x}_{i}$, a separate codebook with $K$ prototypes $\mathbf{C}_{i} = [\mathbf{c}_{i,1}, \mathbf{c}_{i,2}, \ldots, \mathbf{c}_{i,K}]$ is prepared in advance. Subsequently, each subvector $\mathbf{x}_{i}$ is mapped to the corresponding prototype by using an encoding function, $enc_i(\mathbf{x}_i)$, yielding the quantized representation of $\mathbf{x}$ as follows:
\begin{equation}
\label{eq:product_quantization}
    % \mathbf{q}(\mathbf{x}) = [\mathbf{c}_{1,q_{1}(\mathbf{x}_1)}, \mathbf{c}_{2,q_{2}(\mathbf{x}_2)}, \ldots, \mathbf{c}_{M,q_{M}(\mathbf{x}_M)}]
    \mathbf{q}(\mathbf{x}) = [enc_1(\mathbf{x}_1), enc_2(\mathbf{x}_2), \ldots, enc_M(\mathbf{x}_M)]
\end{equation}
The encoding function $enc$ should be carefully designed as it influences the accuracy and efficiency of PQ. Specifically, it is crucial to design $enc$ that has a low computational cost and does not degrade the inference accuracy of neural networks.

%\subsection{Approximate matrix multiplication using product quantization}
\subsection{MADDNESS Algorithm}

\begin{figure}[t!]
    \centering
    \includegraphics[width=0.95\linewidth]{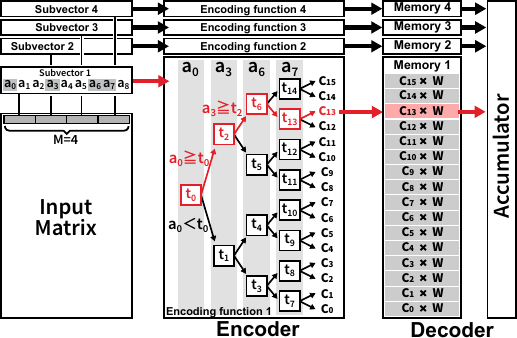}
    \caption{Overview of the MADDNESS algorithm. The input vector is quantized into 16 prototypes by the decision tree classifier, and the corresponding multiplication results are loaded from memory to the accumulator.}
    \label{fig:maddness}
\end{figure}

MADDNESS consists of two primary operations: encoding and decoding. Encoding corresponds to PQ, where the input is divided into subvectors, and the nearest prototype vector is identified for each subvector. In circuit implementations, encoding is equivalent to generating a one-hot LUT address. Decoding involves retrieving and accumulating the precomputed multiplication results of the subvectors and synaptic weights from memory based on the generated address.

Recent studies have proposed several MADDNESS-inspired algorithms. The main difference is the encoding function. The original MADDNESS proposed a balanced binary decision tree (BDT) based encoding function for fast and light inference. PECAN\cite{pecan2022} employs Manhattan distance for encoding, while LUT-NN\cite{lutnet2023} utilizes Euclidean distance, to improve inference accuracy. Stella Nera\cite{stellanera} uses a matrix representation of the BDT to enable MADDNESS to be trained with back-propagation. In the remainder of this paper, the circuit implementations of these algorithms are collectively referred to as MADDNESS-based accelerators.

\subsection{MADDNESS-based DNN Accelerator}

The principal computational operation in MADDNESS accelerators is the calculation of an encoding function $enc$. 
Consequently, there has been a recent focus on the development of circuits capable of executing $enc$ at high speeds with minimal energy consumption. 

\cite{fuketa2023} proposed a method to efficiently compute $enc$ as the similarity between the prototype and the input vector using time-domain analog circuit. 
The operational principle is as follows: firstly, the 6-bit precision input $\mathbf{x}$ and the 6-bit precision prototype $\mathbf{C}_i$ are converted into 60-bit thermometer codes. 
Subsequently, the circuit digital-to-time converter (DTC) is employed to calculate $enc$, whereby the Manhattan distance is converted into signal propagation delay. 
The DTC is constituted by a series of delay chains, with the $k$-th chain being responsible for computing similarity between input $\mathbf{x}$ and $k$-th prototype $\mathbf{C_k}$. The delay chain is constituted by delay cells, comprising variable delay elements and flip-flops connected in series. Each flip-flop pre-stores the prototype converted into thermometer codes. Therefore, by identifying the delay chain that provides the fastest signal propagation delay, it is possible to compute $enc$ in an efficient manner. While this circuit demonstrates high energy efficiency through analog computing, the expansion of the codebook into thermometer codes necessitates a considerable allocation of area and memory capacity, with each $n$-bit codebook requiring $2^n$-bit cells. Furthermore, given the susceptibility of analog computing to PVT variations, the necessity arises for an additional circuit for post-fabrication calibration, with the objective of compensating for the computational errors that are caused by process variations.

\begin{figure}[t!]
    \centering
    \includegraphics[width=0.98\linewidth]{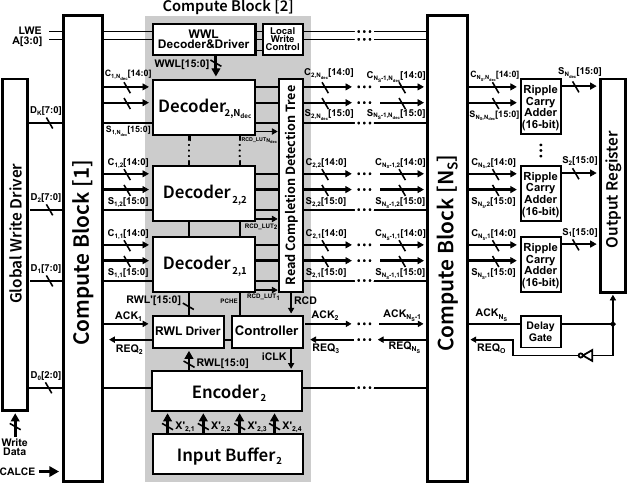}
    \caption{Overall circuit of the proposed LUT-based DNN accelerator.}
    \label{fig:overall}
\end{figure}

To make the accelerator robust to PVT variation, \cite{stellanera} proposed a fully synthesizable all-digital accelerator featuring a decision tree-based classifier for PQ. 
%To efficiently compute $enc$ in digital circuits, the authors propose implementing $enc$ using a binary decision tree. 
Fig.~\ref{fig:maddness} illustrates an example of the decision tree-based $enc$ computation. 
A subvector (9-dimensional in the figure) extracted from the activation vector is classified into the corresponding prototype using a BDT. 
In this process, the elements of the subvector are compared with thresholds to determine the branch to be taken. 
In the example shown, the four elements, $a_0$, $a_3$, $a_6$, and $a_7$, are sequentially compared with the corresponding thresholds from left to right. 
Since there are four elements being compared in this example, the subvector is possible to be classified into $2^4=16$ prototypes. 
Based on the classification results, a lookup table is referenced to read out the precomputed inner product of the prototype and synaptic weights. 
This process is repeated for all four subvectors, and the results are accumulated to perform matrix multiplication.
% The digital implementation allows for performance improvements through advanced process nodes. Furthermore, to reduce the computational load associated with the calculation of $q(\mathbf{x})$, the authors proposed the substitution of a similarity calculation between input $\mathbf{x}$ and a prototype $\mathbf{C}_k$ with a simple decision tree-based classifier. 
It has been reported that scaling the process node from 14nm to 3nm could achieve $4\times$ enhancement in energy efficiency and $32\times$ enhancement in area efficiency. 
However, the standard cell memory-based LUT consumes significantly higher energy and area than the standard SRAM-based LUT. 
Furthermore, the decision tree classifier requires additional memory costs for storing and reading threshold data, resulting in low energy efficiency.

This paper aims to improve energy efficiency by reducing the power consumption of both the PQ circuit and the LUT.
The proposed all-digital macro incorporates a digital-CIM-based decision tree classifier for PQ and LUT based on area-efficient 10T-SRAM with an independent read port, achieving substantial reductions in energy consumption. 
Furthermore, since the proposed macro reliably operates in various PVT conditions by a global clock-free pipeline architecture, it can easily be implemented in advanced process nodes.

\section{Proposed LUT-based DNN Accelerator}
\label{chap:proposed}
\subsection{Overall Architecture}

Fig.~\ref{fig:overall} illustrates the overall macro of the proposed accelerator. The proposed macro consists of multiple serially connected compute blocks, a global write driver to update the contents of SRAM cells in the compute blocks, a 16-bit ripple carry adder (RCA), and an output register. Each compute block comprises an encoder that maps input vectors to the corresponding prototypes and multiple decoders that read precomputed inner product from SRAM based on the output of the encoder and perform accumulation, a self-synchronous pipeline controller for inter-block operations, a read wordline (RWL) driver, a read completion detection (RCD) tree for handshake control, an input buffer, a local write driver, a write wordline (WWL) decoder, and a WWL driver. The encoder includes a 4-level decision tree classifier to perform PQ on the input, while the decoder contains a 10T-SRAM array to store precomputed results, a 16-bit carry-save adder (CSA) for LUT output accumulation, and registers to hold the CSA results.

\begin{figure}[t!]
    \centering
    \includegraphics[width=0.98\linewidth]{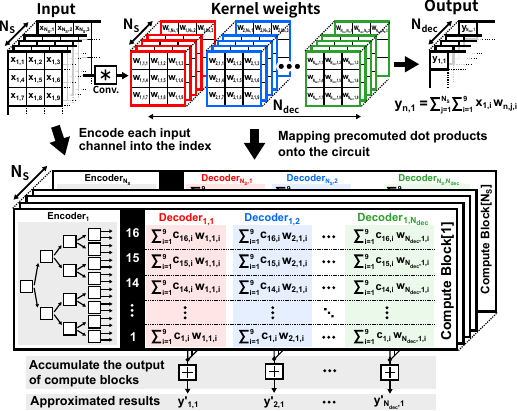}
    \caption{CNN mapping onto the proposed circuit.}
    \label{fig:mapping}
\end{figure}

\begin{figure*}[t!]
    \centering
    \includegraphics[width=0.95\linewidth]{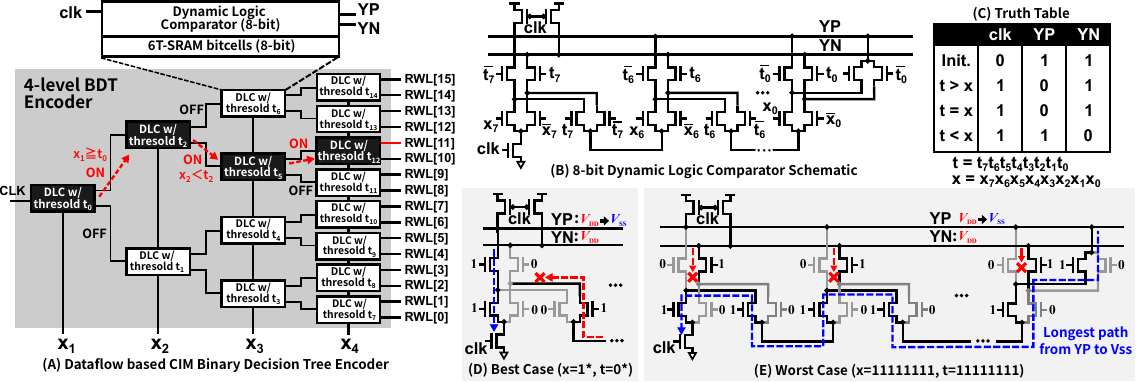}
    \caption{Proposed decision tree-based encoder. (A) Overall schematic of the proposed encoder circuit. (B), (C) Schematic and truth table of the proposed dynamic logic comparator. (D), (E) Best and worst case operation of the DLC.}
    \label{fig:bdt}
\end{figure*}

The proposed circuit has two adjustable parameters, $N_\textrm{S}$ and $N_\textrm{dec}$. $N_\textrm{S}$ is the number of pipeline stages, and $N_\textrm{dec}$ is the number of decoders per compute block. In convolutional neural network\cite{cnn1998} inference, $N_\textrm{S}$ corresponds to the number of input channels that can be processed concurrently, while $N_\textrm{dec}$ corresponds to the number of weight kernels that can be processed simultaneously, as shown in Fig.~\ref{fig:mapping}. Increasing either parameter can reduce the overhead of peripheral circuits per operation. Notably, increasing  $N_\textrm{dec}$ reduces the encoder overhead to $1/N_\textrm{dec}$, thereby enhancing area and energy efficiency. However, increasing $N_\textrm{dec}$ also raises the WL wiring resistance within the LUT and the depth of the RCD tree, leading to increased latency. Additionally, since an internal critical path of each block limits block latency, worst-case delays may be increased for larger circuits. The optimal value for $N_\textrm{dec}$ is evaluated in Section ~\ref{chap:evaluation}.

The operation of the compute block is as follows. First, prior to the inference, the precomputed dot products of the prototypes and synaptic weights are loaded in the corresponding SRAM cells embedded in decoders. In this paper, we employed an 8-bit integer precision. During the inference, the self-synchronous controller sets the block to an operation mode, which then activates the encoder to convert the subvector into the corresponding prototype. Similarly to \cite{stellanera}, the proposed circuit employed a BDT for the encoding function, $enc$. Based on the mapped prototype, the decoder retrieves the precomputed dot product from the SRAM array. The retrieved data is accumulated with the output of the preceding block by the CSA, stored in the output latch of each decoder, and transmitted to the CSA of the subsequent block. Once all the lookup operations have been completed, the final summation is computed using the RCAs, and the result is stored in the output register.

Notably, all operations in the proposed circuit are performed asynchronously.
%It should be noted that all operations are performed asynchronously. 
In a typical clock-synchronized pipeline, the longest critical path among all stages determines the latency~\cite{pipeline1977}. 
However, the asynchronous pipeline architecture allows each block to process the next input immediately, without waiting for the clock's rising edge, thereby minimizing latency~\cite{asynchronous1995}. 
Furthermore, this feature dynamically determines the operating speed, ensuring stable operation across various PVT conditions. 
A four-phase handshake protocol\cite{handshake1999} is used for synchronization between compute blocks. 
Additionally, as will be detailed in Sec.~\ref{subsec:encoder}, the BDT inference engine also employs dynamic circuits, allowing only the necessary comparators to be activated automatically, which significantly reduces power consumption.

\subsection{Decision Tree-based Encoder}
\label{subsec:encoder}

As mentioned above, the proposed circuit uses a BDT in the encoder to classify subvectors into prototypes. Fig.~\ref{fig:bdt}(A) illustrates a block diagram of the encoder part of the proposed circuit. As shown, the encoder consists of dynamic logic comparators (DLCs) that compare the threshold values $\mathbf{t}$ with the elements of the subvector $\mathbf{x}$. To classify the subvectors into one of 16 prototypes, the proposed encoder connects 15 DLCs in a tournament structure. 
% By employing dynamic logic in the comparators, the encoder circuit achieves faster computation and self-synchronous operation. 
Conventional clock-synchronous-based circuit proposed in \cite{stellanera} require a global clock and many internal registers, resulting in high-power consumption.
In contrast, the proposed encoder circuit achieves low-power computation through self-synchronous operation, eliminating both global clock and internal registers by employing dual-rail dynamic logic in the comparators.
The red dotted line in Fig.~\ref{fig:bdt}(A) shows an example of how a subvector is classified into a prototype.
% On the layout, they are placed in a 3$\times$5 rectangle. 
First, all DLCs are precharged. Note here that each DLC stores a pre-learned threshold. Then, the root DLC is activated to compare the threshold $t_0$ with an element of the subvector $\mathbf{x_1}$. Depending on the result of the comparison, the next level DLC is selected to activate. By activating only the necessary DLCs, the circuit minimizes the discharge of precharged charges, thereby improving energy efficiency.
% Inputs are sequentially compared with thresholds from the root node downwards, ultimately outputting a one-hot WL signal. Leveraging the dynamic circuit property where computation is only executed upon clock input, only one DLC is selectively activated in each layer. This approach minimizes dynamic power by preventing the activation of non-contributory DLCs in the output generation.

% \subsubsection{Dynamic logic comparator for low power BDT-search}
Fig.~\ref{fig:bdt}(B) and (C) show the schematic and the truth table of the DLC composed of a dual-rail dynamic logic. The DLC consists of eight 1-bit dynamic comparators connected in series. First, during the precharge phase (clk=0), both output nodes, YP and YN, are precharged to VDD. Additionally, the values of the two operands are set up in this phase. Then, in the evaluation phase (clk=1), the PMOS turns off, and depending on the input and threshold values, YP or YN is pulled down to VSS through the NMOS footer, as shown in the truth table.
Note here that each 1-bit comparator discharges either YP or YN based on the comparison result if the comparison can be determined by the digit it is responsible for. If the comparison cannot be determined by its digit alone, the connection to the 1-bit comparator of the lower digit is activated, allowing YP/YN to be discharged based on the comparison of the lower digit. 
% In the proposed DLC, if the difference between the input and the threshold is closer to the MSB, the number of transistors between the YP/YN nodes and VSS is reduced, enabling faster operation.
This method not only reduces the number of transistors required compared to comparators using naive subtractors, but also speeds up the comparison operation by immediately completing the calculation for those that can be determined by the higher digits alone, as shown in Figs.~\ref{fig:bdt}(D) and (E).

\subsection{Two-port 10T-SRAM-based Decoder with CSA}

\begin{figure}[t!]
    \centering
    \includegraphics[width=0.9\linewidth]{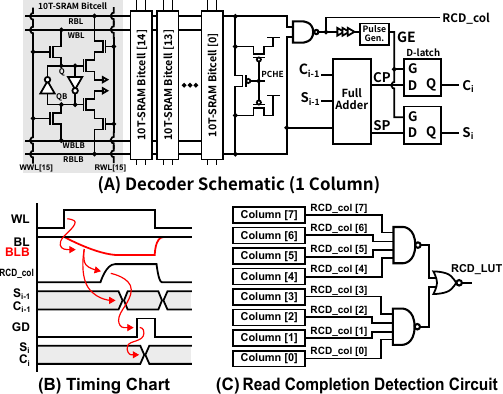}
    \caption{10T-SRAM based decoder with CSA unit. (A) Schematic of one SRAM column, (B) Timing chart of the decoding operation, (C) Schematic of the RCD circuit.}
    \label{fig:decoder}
\end{figure}

% 回路構成の説明をマクロのところでするか，ここでするか
After mapping the subvector to the corresponding prototype with the encoder, decoders retrieve the precomputed dot product between the prototype and the synaptic weights from SRAM and perform accumulation. 
%The schematic of the proposed decoder is shown in Fig.~\ref{fig:decoder}. 
The proposed decoder consists of a 16-row, 8-column two-port 10T-SRAM array, a 16-bit CSA for accumulation, an output latch, and a 2NAND-1NOR circuit for RCD. This design facilitates high-speed data readout without the need for a sense amplifier (SA), as demonstrated in~\cite{tsmc2024}.

Fig.~\ref{fig:decoder}(A) and (B) illustrate the schematic and timing chart of a single column. The read operation proceeds as follows: first, based on the encoder output, one RWL in the column is asserted. Depending on the stored value, the selected SRAM cell fully discharges either RBL or RBLB. Subsequently, the full adder (FA) computes the carry and sum, and the output of the NAND gate (RCD\_col) transitions to high. After a brief delay, the GE signal activates the latch, holding the FA outputs. When all RCD\_col[i] signals in the LUT output high, they are combined into an overall RCD\_LUT[i] signal via a NAND-NOR tournament, as depicted in Fig.~\ref{fig:decoder}(C). All RCD\_LUT[i] signals in the same block are also aggregated into a single RCD[i] signal, as shown in Fig.~\ref{fig:overall}, for the handshake control.

The proposed SRAM array incorporates a column-level RCD function to make the decoder robust to PVT variation. Traditional methods of estimating read latency often use a replica SRAM column, which accurately estimates read speed by mimicking the WL and BL layouts and conditions~\cite{replica1998}. However, this approach is prone to column-to-column variability, requiring additional delay circuits for adjustment. In contrast, the proposed design features an independent RCD circuit for each column, enabling accurate detection even under high variability conditions. In addition, latch timing for the CSA outputs is generated simultaneously, effectively preventing setup violations over a wide range of PVT conditions.

\section{Evaluation}
\label{chap:evaluation}

The performance of the proposed circuit was evaluated via post-layout simulation using Synopsys HSPICE\textregistered, based on a layout implemented in a commercial 22nm bulk-CMOS process. The total chip area and core area of the macro were 0.66\,\mm\,and 0.20\,\mm, including 64kb SRAM ($N_\textrm{dec}=16$, $N_{\textrm{S}}=32$).

\begin{figure}[t!]
    \centering
    \includegraphics[width=0.98\linewidth]{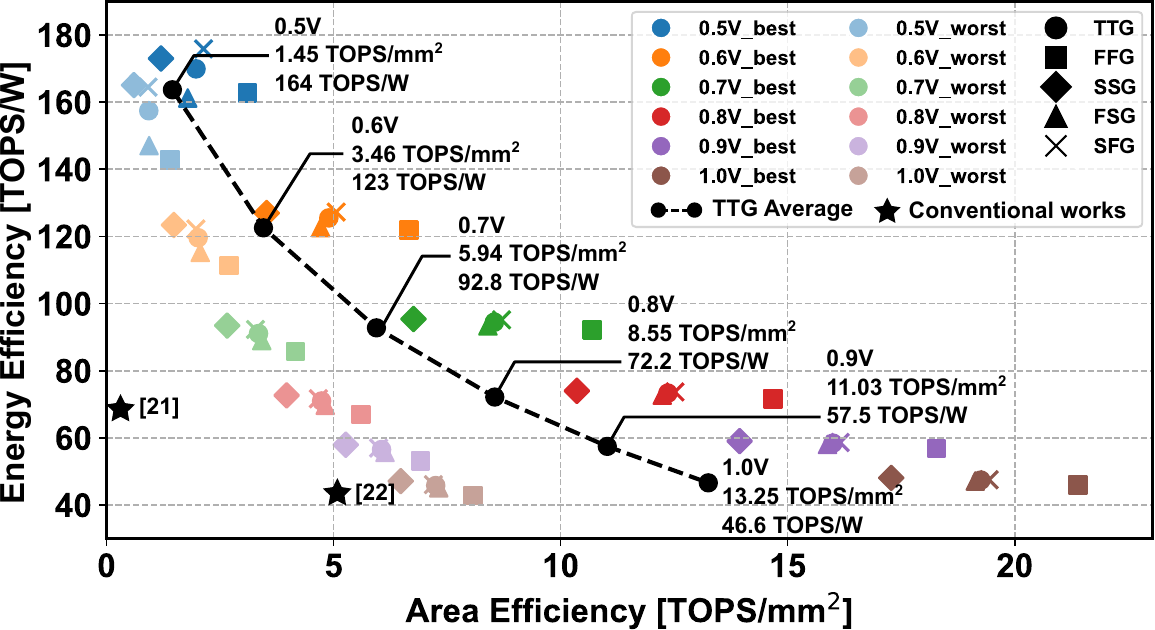}
    \caption{Energy and area efficiency of the proposed circuit across various supply voltages and process corners. The black dashed line shows the average performance of the worst and best case at TTG corner. The upper right edge indicates better performance.}
    \label{fig:topspw_vs_freq}
\end{figure}

Fig.~\ref{fig:topspw_vs_freq} evaluates the performance of the proposed circuit in terms of energy efficiency (TOPS/W) and area efficiency (TOPS/\mm) across various supply voltages (0.5V-1.0V) and process corners  (TTG, FFG, SSG, SFG, FSG).  Simulation settings are 25\cer, $N_\textrm{dec}=4$, and $N_\textrm{S}=4$. 
Since the latency of the BDT encoder varies depending on the input and BDT threshold values, as described in Sec.~\ref{subsec:encoder}, both the worst and best cases are shown in the figure. The black dashed line illustrates the average of the worst and best performances at the TTG corner.
At 0.5V, the circuit achieves the highest energy efficiency of 164 TOPS/W, while at 1.0V, it demonstrates the highest area efficiency of 13.25 TOPS/\mm on average, highlighting a trade-off between energy consumption and throughput.  
%The supply voltage determined the energy efficiency almost exclusively, with the SSG corner having a lower throughput than the FFG corner but similar energy efficiency.
The energy efficiency of the proposed circuit is determined mainly by supply voltage, which is nearly constant regardless of process corners and BDT encoder latency.
Compared to conventional works (marked by black stars), the proposed circuit outperforms in both energy and area efficiency, particularly excelling in low-power scenarios.

\begin{figure}[t!]
    \centering
    \includegraphics[width=0.95\linewidth]{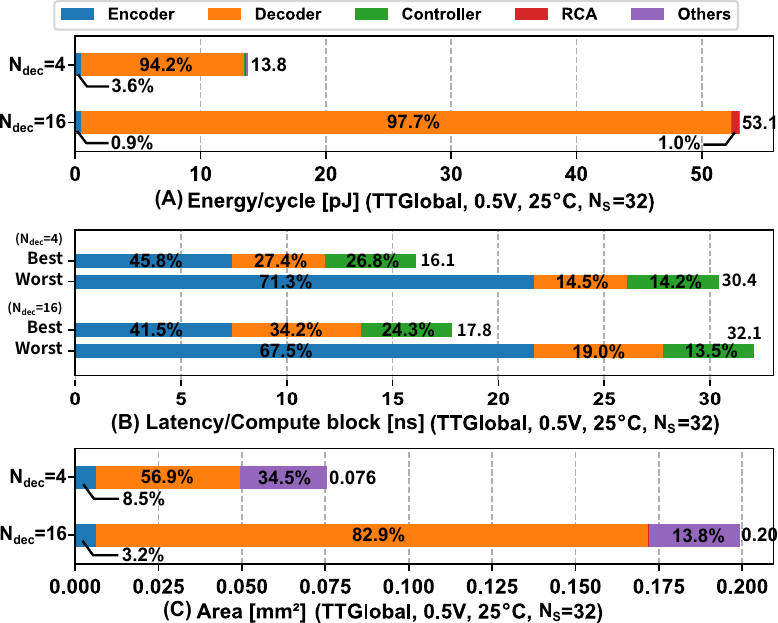}
    \caption{(A) Energy, (B) latency, and (C) area breakdown based on post-layout simulation results.}
    \label{fig:ppa}
\end{figure}

Fig.~\ref{fig:ppa} shows the energy, latency, and area breakdown for $N_\textrm{dec} = 4, 16$ and $N_\textrm{S} = 32$ at 0.5V. For energy, over 94\,\% of consumption for both $N_\textrm{dec} = 4$ and  $N_\textrm{dec} = 16$  was attributed to the decoder, including the SRAM array and CSA. Similarly, the decoder accounted for a significant proportion (50\,\% to 80\,\%) of the area. This is because the energy and area of the decoder increase as $\mathcal{O}(N_\textrm{dec} \cdot N_\textrm{S})$, while other components increase only as $\mathcal{O}(N_\textrm{dec})$ or $\mathcal{O}(N_\textrm{S})$. In contrast, for latency, the encoder constituted the largest proportion over 40\,\% to 70\,\%. Comparing $N_\textrm{dec} = 4$ and $N_\textrm{dec} = 16$, the relative proportion of non-decoder components for energy and area decreases. On the other hand, the decoder's latency increased slightly due to the increased delay in the RCD.

\begin{table}[t!]
\begin{threeparttable}[t]
\centering
\scriptsize
\caption{Performance for different $N_\textrm{dec}$}
\begin{tabular}{c|c|c|c|c}
\hline
\textbf{Voltage} & \textbf{$N_\textrm{dec} = 4$} & \textbf{$N_\textrm{dec} = 8$} & \textbf{$N_\textrm{dec} = 16$} & \textbf{$N_\textrm{dec} = 32$} \\ \hline
\multicolumn{5}{c}{\textbf{Energy efficiency [TOPS/W]}} \\ \hline
0.5V & 167.5 & 171.8 ($+$2.6\%) & 174.0 ($+$3.9\%) & 174.9 ($+$4.4\%) \\ \hline
0.8V & 73.0 & 74.4 ($+$1.0\%) & 75.1 ($+$1.0\%) & 75.4 ($+$1.0\%) \\ \hline
\multicolumn{5}{c}{\textbf{Area efficiency [TOPS/\mm]}} \\ \hline
0.5V & 1.4 & 1.8 ($+$28.6\%) & 2.0 ($+$42.9\%) & 2.0 ($+$42.9\%) \\ \hline
0.8V & 8.7 & 10.8 ($+$24.1\%) & 11.3 ($+$29.9\%) & 11.5 ($+$32.2\%) \\ \hline
\end{tabular}
\begin{tablenotes}
\item[*] The numbers in parentheses indicate the improvement rate from $N_\textrm{dec} = 4$.
\end{tablenotes}
\label{tab:ndec_performance}
\end{threeparttable}
\end{table}

To estimate the impact of $N_\textrm{dec}$ values on performance more precisely, Table~\ref{tab:ndec_performance} shows the area and energy efficiency for $N_\textrm{dec} =$ 4, 8, 16,  and  32. Simulations were conducted under TTG corner at 25\cer, $N_\textrm{S}=32$, and two supply voltage levels: 0.5 V for maximum energy efficiency and 0.8 V, the nominal voltage of 22nm process.
The figure indicates increasing $N_\textrm{dec}$ improves both area and energy efficiency, with $N_\textrm{dec} = 32$ demonstrate 32.2\,\% to 42.9\,\% improvements in area efficiency and 1.0\,\% to 4.4\,\% improvements in energy efficiency compared to $N_\textrm{dec} = 4$. However, further increases in $N_\textrm{dec}$ do not yield additional performance enhancements. For instance, the performance gain between $N_\textrm{dec} = 32$ and  $N_\textrm{dec} = 16$ is 0\,\% to 2\,\%, which is almost negligible. 
%Since larger $N_\textrm{dec}$ values make the circuit vulnerable to local variations, it is recommended that $N_\textrm{dec} = 16$ to achieve an optimal balance between performance and reliability.
Since larger $N_\textrm{dec}$ values make the circuit vulnerable to local variations, $N_\textrm{dec} = 16$ is recommended to achieve an optimal balance between performance and reliability.

\begin{table}[t!]
\centering
\begin{threeparttable}[t!]
    \centering
    \scriptsize
    \caption{Comparison to prior accelerators.}
    \setlength{\tabcolsep}{4pt}
    \begin{tabular}{c|c|c|c|c} \hline 
                                                        & {TCAS-I$^{\prime}$23~\cite{fuketa2023}} &  {arXiv$^{\prime}$23~\cite{stellanera}} & \multicolumn{2}{c}{\tablenl{Proposed\\($N_\textrm{dec}$=16, $N_\textrm{S}$=32)}}            \\ \hline 
        \tablenl{Measured\\/Simulated}                  &  Measured                     &  Simulated                    & \multicolumn{2}{c}{Simulated}                                    \\ \hline 
        Operation Mode                                  &  \tablenl{MADDNESS\\(Analog)} &  \tablenl{MADDNESS\\(Digital)}& \multicolumn{2}{c}{\tablenl{MADDNESS\\(Digital)}}                \\ \hline 
        Process [nm]                                    &  65 (Planar)                           &  14 (FinFET)                           & \multicolumn{2}{c}{22 (Planar)}                                           \\ \hline 
        Power Supply [V]                                &  0.35/0.6/1.0\tnote{(2)}                 &  0.55                         & 0.5                               & 0.8\tnote{(1)}                           \\ \hline 
        Area [\mm]                                      &  0.31                         &  0.57                         & \multicolumn{2}{c}{0.20}                                          \\ \hline 
        Frequency [MHz]                                 &  77                           &  624                          & 31.2-56.2                         & 144-353                       \\ \hline 
        LUT Precision                                   &  INT8\tnote{(3)}                         &  INT8                         & \multicolumn{2}{c}{INT8}                                         \\ \hline 
        \tablenl{Throughput\\{[TOPS]}}                  &  0.089                        &  2.9                          & 0.28-0.51                         & 1.33-3.26                     \\ \hline 
        \tablenl{Energy Efficiency\\{[TOPS/W]}}         &  69                           &  43.1                         & 174                               & 75.1                          \\ \hline 
        \tablenl{Area Efficiency \\ {[TOPS/\mm]}}       & \tablenl{0.29\\(0.40)\tnote{(4)}}        & \tablenl{5.1\\(2.70)\tnote{(4)}}         & 2.01                              & 11.34                      \\ \hline 
        \tablenl{ResNet9 Acc.\\(CIFAR-10)}              & 89.0                          & 92.6                          & 92.6                              & 92.6                          \\ \hline 
        %\tablenl{Energy/op. {[fJ]}\\(Encoder)}          & 7.47                          & 20.25/$N_\textrm{dec}$        & 0.86/$N_\textrm{dec}$             & 1.72/$N_\textrm{dec}$         \\ \hline 
        \tablenl{Energy/op. {[fJ]}\\(Encoder)}          & 7.47                          & 1.27                          & 0.054                             & 0.11                          \\ \hline 
        \tablenl{Energy/op. {[fJ]}\\(Decoder)}          & 7.02\tnote{(5)}                          & 16.47                         & 5.6                               & 14.7                          \\ \hline 
    \end{tabular}
    \begin{tablenotes}
        \item {$^1$}\,Nominal voltage of the 22nm process.  {$^2$}\,Multiple $V_\textrm{DD}$ structure.
        \item {$^{3}$}\,Adjustable between INT4 and INT32. {$^{4}$}\,Scaled to 22nm process.
        \item  {$^{5}$}\,Accumulator is not included.
    \end{tablenotes}
    \label{tab:comparison_table}
\end{threeparttable}
\end{table}

Finally, Table~\ref{tab:comparison_table} provides a comparison of the proposed circuit ($N_\textrm{dec}=16$, $N_\textrm{S}=32$) with state-of-the-art works. 
Area efficiency was normalized to 22nm process for a fair comparison. For example, circuits implemented in a 65nm process were scaled by $(65/22)^2$ based on the reported data. However, since \cite{fuketa2023} employs analog computation in the encoder, area scaling was applied only to the digital parts. 
The simulation results show that the proposed circuit achieved 2.5$\times$ higher energy efficiency and 5$\times$ higher area efficiency at a supply voltage of 0.5V compared to ~\cite{fuketa2023}. Note that the performance of \cite{fuketa2023} does not include accumulator overhead. 
Compared to \cite{stellanera},  the proposed circuit achieves 4.0$\times$ the energy efficiency, but the area efficiency is 25\,\% lower. 
However, at a supply voltage of 0.8 V, the proposed circuit achieved 1.7$\times$ increase in the energy efficiency and 4.2$\times$ increase in the area efficiency, exceeding in both metrics.
%At a supply voltage of 0.8V, the proposed circuit achieved 1.5$\times$ increase in the energy efficiency and 4.2$\times$ increase in the area efficiency compared to ~\cite{stellanera}. 
Instead of high energy efficiency, the proposed circuit has a lower operating frequency compared to the \cite{stellanera}, resulting in lower throughput.
However, the smaller LUT layout of the proposed circuit allows it to maintain comparable area efficiency.
%the proposed circuit is comparable to the \cite{stellanera} in terms of area efficiency due to the smaller layout of the LUT. 
Therefore, the proposed circuit can be applicable to situations where higher throughput is required by dividing the macros, although an additional adder is required.
The performance improvement in energy efficiency is mainly attributed to the energy reduction in the decoder by employing an efficient 10T-SRAM array, which consumes 66\,\% less energy than standard cell memory-based LUT~\cite{stellanera}. 
Additionally, the encoder reduced energy consumption by 95\,\% owing to eliminating the threshold readout cost and internal registers.
It is also noteworthy that the classification accuracy on the CIFAR-10 dataset using ResNet9 was identical to that reported in \cite{stellanera}.

\section{Conclusion}
\label{chap:conclusion}
This paper proposed a novel MADDNESS-based all-digital accelerator that significantly enhances energy and area efficiency for deep neural network inference. By employing a self-synchronous pipeline architecture, the design eliminates the need for a global clock, ensuring robust operation across various PVT conditions. The integration of a low-power binary decision tree encoder enables efficient PQ, while the implementation of an area-efficient lookup table using two-port 10T-SRAM substantially reduced energy consumption during decoding. Post-layout simulations using a commercial 22nm process demonstrated a 2.5$\times$ improvement in energy efficiency (174 TOPS/W) and a 5$\times$ increase in area efficiency (2.01 TOPS/\mm) compared to conventional accelerators.

\section*{Acknowledgement}
This work was supported by Japan Science and Technology Agency (JST), PRESTO Grant Number JPMJPR22B1 and  JST BOOST, Grant Number JPMJBS2407, Japan.

%IEEE conference templates contain guidance text for composing and formatting conference papers. Please ensure that all template text is removed from your conference paper prior to submission to the conference. Failure to remove the template text from your paper may result in your paper not being published.

\newpage
%引用を整理する．
% \bibliographystyle{IEEEtran}
% \bibliography{ref}
\printbibliography

\end{document}